# Short Report: Study variability in recent human challenge experiments with *Plasmodium falciparum* sporozoites ('PfSPZ Challenge')

Andrew A. Lover[1]*

*Malaria Elimination Initiative*
*University of California, San Francisco*
*San Francisco, California*
*Email: Andrew.Lover@ucsf.edu or Andrew.A.Lover@gmail.com*

**Abstract**

There has been renewed interest in the use of sporozoite-based approaches for malaria vaccination and controlled human infections, and several sets of human challenge studies have recently completed. A study undertaken in Tanzania and published in 2014 found dose-dependence between 10,000 and 25,000 sporozoite doses, as well as divergent times-to-parasitemia relative to earlier studies in European volunteers. However, this analysis shows that these conclusions are based upon suboptimal analytical methods; with more optimal analysis, there is no evidence for dose-dependence within this dose range; and more importantly, no evidence for differences in event times between Dutch and Tanzanian study sites. While these finding do not impact the reported safety and tolerability of PfSPZ, they highlight critical issues that should be comprehensively considered in future challenge studies.

Experimental injection of infective malaria parasites as sporozoites is currently undergoing a re-evaluation as a viable vaccine strategy for prevention of malaria infections in humans, and for controlled human malaria infection (CHMI). A recently published paper reports on the use of cryopreserved *Plasmodium falciparum* sporozoites ('PfSPZ Challenge,' Sanaria Inc.) for challenge experiments in human volunteers in Tanzania.[1]

I applaud the authors' efforts to bring these critically important studies to malaria-endemic settings in sub-Saharan Africa, and agree that this is truly a milestone in global efforts towards development of a sporozoite vaccine against malaria and controlled human infections. However, in light of the enormous financial, technical and ethical challenges underlying these bridging studies I believe it is critical that they provide accurate guidance for future trials.

There are several aspects of the published analysis that have major impacts on Shekalaghe *et al*'s conclusions. With more appropriate analysis, and contrary to the reported results, their data do not provide any evidence for dose-response in sporozoite dosing, and more importantly there is no evidence for statistically significant differences in reported times-to-parasitemia between Tanzanian and Dutch cohorts.

First and foremost, the prepatent periods (time from sporozoite exposure to febrile illness) in this study have been compared as geometric means via a non-parametric Wilcoxon rank-sum test. While this is common practice in some fields, there are more appropriate methods for the analysis of time-to-event data,[2] and a large body of literature exists with approaches that allow comprehensive analyses of these types of data.[3] In earlier work with *Plasmodium vivax*, it was shown that application of this, and other suboptimal methods may lead to erroneous conclusions regarding dose-dependence in sporozoite inoculations.[4]

A second major issue, which has received extensive attention in the epidemiological and clinical trials literature but was not considered by the authors, is one of potential biases from exclusion in the analysis of any patients originally randomized.[5] This problem is especially acute in these studies due to the limited sample size. In the published earlier work three patients were removed from the analysis- however these data need to be included in an 'intent-to-treat' analysis to avoid potentially biased conclusions. The 'failed' infections that were treated but removed from the analysis may well have been delayed incubation periods- while this phenomenon is better documented in *P. vivax* infections, it has also been reported in *P. falciparum* infections.[6,7] While extended follow-up is not feasible or ethical in these challenge studies, a censoring date should be explicit in the research protocol.

To quantify the impact of these issues, I have re-analyzed these study data using statistical analyses optimized for these types of data (see Figure 1 and Table 1). This re-analysis consisted of 1) unadjusted comparisons using logrank-type tests (using alternatives where the survival curves are crossing) followed by 2) multivariate models.

While the authors' report that "Volunteers in the 10,000 PfSPZ group had a significantly different pre-patent period than in the 25,000 PfSPZ group (geometric mean [GM] of 15.4 and 13.5 days, Wilcoxon, P = 0.023)"[1] I find no evidence of differences using a logrank test (p= 0.179), and no differences in qPCR times (Renyi logrank p = 0.266). Kaplan-Meier plots of these results are show in Figure 1, Panels A and B. Moreover, the potential impact of not fully considering patient censoring is readily apparent in the differences between the full and reduced datasets (Table 1).

Additionally, the authors when comparing results to their earlier studies[8] were surprised to find that "...the GM pre-patent period for the Tanzanians who received the 10,000 PfSPZ dosage regimen was 15.4 days, and the GM pre-patent period for the Dutch was 12.6 days ($P = 0.0192$,

Wilcoxon 2-tailed)." Unadjusted comparisons of these times (N=18) taking into account time-to-event data provides no evidence of differences (logrank p= 0.288; Renyi logrank = 0.110).

Finally, to comprehensively assess the overall impacts of dose and study site on event times, multivariate Cox and Poisson were used to provide full covariate adjustment, and show no evidence for statistically significant differences in event times between any of the dose categories or between the two study sites (Figure 1, panel C; results by microscopy, Table 2; results by PCR, Table 3) in a direct comparison amongst the combined cohorts (N=42).

Importantly, while the results from this re-analysis do not impact the reported safety or tolerability of PfSPZ challenge, they highlight critically important areas that require further detailed consideration. First, these results suggest that there is no evidence for dose response within this range of sporozoite inoculations, which consequent potential to complicate dose ranging in future studies. Secondly, these results should allay the authors' concerns regarding any potential differences in response to challenge between Dutch and Tanzania volunteers in future work, and, obviate the need for the hypotheses suggested in their discussion (e.g. genetic differences in merozoites release).

In summary, greater consideration and wider utilization of comprehensive survival analyses to provide sound conclusions towards informing further work using these challenging and critically important experiments.

Note: Appendix 1 contains detailed statistical methods; Appendix 2 contains the dataset for this analysis.

**Funding statement**
No specific funding was used in this work.

**Conflict of interest statement**
I declare I have no conflicts of interest to report.

**Figure**

Figure 1. Kaplan-Meier curves comparing times-to-parasitemia by sporozoite dose groups in human sporozoite challenge studies. *Panel A*. Tanzanian cohort via microscopy; *Panel B*. Tanzanian cohort via PCR; *Panel C*. Combined Tanzanian and Dutch cohorts via PCR.

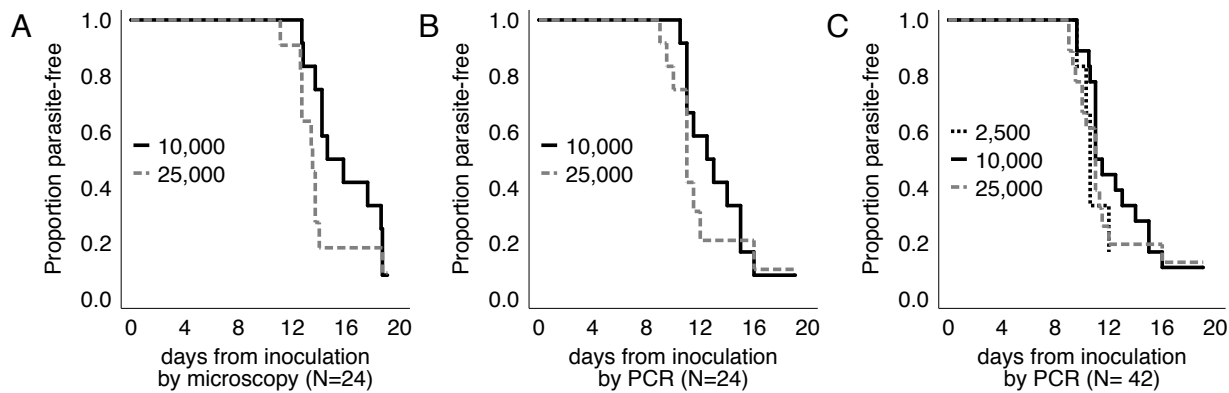

**Tables**

Table 1. Non-parametric comparisons between time-to-parasitemia in the 10,000 vs. 25,000 sporozoite dose groups; full data and truncated data (Tanzanian study).

| Study Endpoint | Comparison | Test | Full data (N=24) p-value | Reduced data (N=21) p-value |
|---|---|---|---|---|
| **Parasitemia (microscopy)** | 10,000 vs. 25,000 dose | logrank | 0.179 | 0.050 |
| **Parasitemia (PCR)** | 10,000 vs. 25,000 dose | Renyi-logrank | 0.266 | 0.102 |

Table 2. Multivariate Cox model for time-to-parasitemia by microscopy in human sporozoite challenge trials, comparing dose and combined study cohorts (N= 42). Note: HR= Hazard ratio.

| Outcome | Risk factor | HR | HR 95% CI | p-value |
|---|---|---|---|---|
| **Parasitemia via microscopy** | 2,500 sporozoite dose | 1.17 | 0.31 – 4.45 | 0.821 |
| | 10,000 sporozoite dose | ref. | - | - |
| | 25,000 sporozoite dose | 1.34 | 0.67 – 2.71 | 0.408 |
| **Study cohort** | Tanzania | ref | - | - |
| | Netherlands | 1.21 | 0.52 – 2.81 | 0.655 |

Table 3. Multivariate Poisson model for time-to-parasitemia by PCR in human sporozoite challenge trials, comparing dose and combined study cohorts (N= 42). Notes: Poisson model utilized due to evidence of non-proportional hazards in Cox models; IRR= Incidence rate ratio.

| Outcome | Risk factor | IRR | IRR 95% CI | p-value |
|---|---|---|---|---|
| **Parasitemia via PCR** | 2,500 sporozoite dose | 0.98 | 0.32 – 3.03 | 0.973 |
| | 10,000 sporozoite dose | ref. | - | - |
| | 25,000 sporozoite dose | 1.02 | 0.51 – 2.07 | 0.950 |
| **Study cohort** | Tanzania | ref | - | - |
| | Netherlands | 1.00 | 0.47 – 2.12 | 0.997 |

**Appendix 1- Detailed Methods**

The endpoints for the survival analyses were taken from table 2 in reference[1] and table 2 in reference.[8] In the Tanzanian studies, two of the excluded patients were censored at the reported times (volunteer 40055-20 at 19 days, and volunteer 50057-20 at 11 days); insufficient data was reported to censor the third volunteer (40010-20) who was also censored on day 19. Patients in the Dutch studies were censored on day 21 when presumptive treatment was given. Dosage was modeled as an ordinal variable; inclusion as a numeric had no impact on model fit.

Model parsimony was assessed using Akaike and Bayesian information criteria; and all Cox models were assessed for proportional hazard violations using scaled Schoenfeld residuals plus graphical comparisons with Kaplan-Meier curves.

In situations where Kaplan-Meier curves cross, conventional logrank-type tests are invalid; therefore Renyi-family tests (developed for this specific situation) were used to compare unadjusted survival between groups.[9,10]

In multivariate Cox models for PCR-endpoints in the combined cohort, proportional hazard assumptions were not met, so Poisson models are presented; other modeling strategies (stratified Cox, flexible parametric, and parametric models) produced comparable estimates and consistent conclusions. Standardized residuals were used to assess fit of multivariate models.

The Renyi tests were implemented using the –survmisc- package;[10] analyses were performed in Stata 13.1 (College Station, Texas, USA), and R software (version 3.0.1).[11]

# Appendix 2- Dataset for analysis

| index | study_id | PP_slide | dose | PP_pcr | failure | study |
|---|---|---|---|---|---|---|
| 1 | 10002-20 | 18.6 | 10000 | 16 | 1 | Tanzanian |
| 2 | 10023-20 | 18.7 | 10000 | 15 | 1 | Tanzanian |
| 3 | 30035-20 | 18.7 | 10000 | 14 | 1 | Tanzanian |
| 4 | 40010-20 | 19 | 10000 | 19 | 0 | Tanzanian |
| 5 | 50041-20 | 14.6 | 10000 | 13 | 1 | Tanzanian |
| 6 | 60008-20 | 12.8 | 10000 | 10.5 | 1 | Tanzanian |
| 7 | 60026-20 | 12.7 | 10000 | 11.5 | 1 | Tanzanian |
| 8 | 70001-20 | 14.2 | 10000 | 11 | 1 | Tanzanian |
| 9 | 70014-20 | 15.8 | 10000 | 12.5 | 1 | Tanzanian |
| 10 | 70031-20 | 14.2 | 10000 | 11 | 1 | Tanzanian |
| 11 | 70044-20 | 17.6 | 10000 | 15 | 1 | Tanzanian |
| 12 | 90047-20 | 13.7 | 10000 | 11 | 1 | Tanzanian |
| 13 | 20056-20 | 18.7 | 25000 | 16 | 1 | Tanzanian |
| 14 | 20064-20 | 11.1 | 25000 | 9 | 1 | Tanzanian |
| 15 | 20070-20 | 12.6 | 25000 | 9.5 | 1 | Tanzanian |
| 16 | 30053-20 | 13.7 | 25000 | 12 | 1 | Tanzanian |
| 17 | 30060-20 | 13.5 | 25000 | 11 | 1 | Tanzanian |
| 18 | 40055-20 | 19 | 25000 | 19 | 0 | Tanzanian |
| 19 | 40068-20 | 13.4 | 25000 | 11 | 1 | Tanzanian |
| 20 | 50050-20 | 12.7 | 25000 | 10 | 1 | Tanzanian |
| 21 | 50057-20 | 11 | 25000 | 11 | 0 | Tanzanian |
| 22 | 60051-20 | 12.7 | 25000 | 11 | 1 | Tanzanian |
| 23 | 60072-20 | 14 | 25000 | 11.5 | 1 | Tanzanian |
| 24 | 80058-20 | 13.7 | 25000 | 11 | 1 | Tanzanian |
| 25 | 696 -18 | 12.3 | 2500 | 9.6 | 1 | Dutch |
| 26 | 711-08 | 14 | 2500 | 12 | 1 | Dutch |
| 27 | 795-06 | 21 | 2500 | 21 | 0 | Dutch |
| 28 | 935-01 | 14 | 2500 | 10.6 | 1 | Dutch |
| 29 | 937-20 | 12.3 | 2500 | 10.6 | 1 | Dutch |
| 30 | 940 -14 | 12.3 | 2500 | 10.3 | 1 | Dutch |
| 31 | 119-03 | 12.6 | 10000 | 9.6 | 1 | Dutch |
| 32 | 603-11 | 13 | 10000 | 11 | 1 | Dutch |
| 33 | 736-04 | 11 | 10000 | 9.6 | 1 | Dutch |
| 34 | 783-25 | 13.3 | 10000 | 10.6 | 1 | Dutch |
| 35 | 788-21 | 14 | 10000 | 11 | 1 | Dutch |
| 36 | 925-26 | 21 | 10000 | 21 | 0 | Dutch |
| 37 | 647-30 | 14 | 25000 | 9.3 | 1 | Dutch |
| 38 | 720-13 | 12.3 | 25000 | 10.3 | 1 | Dutch |
| 39 | 789-15 | 21 | 25000 | 21 | 0 | Dutch |
| 40 | 806-09 | 12.3 | 25000 | 9 | 1 | Dutch |
| 41 | 909-29 | 14.3 | 25000 | 11.3 | 1 | Dutch |
| 42 | 926-24 | 12.3 | 25000 | 10 | 1 | Dutch |

Codebook

    index: ID for analysis

    study_id: patient ID in original publications

    PP_slide: reported prepatent period via microscopy (days)

    dose: reported sporozoite dosage

    PP_pcr: reported prepatent period via PCR (days)

    failure: 1= parasitemic; 0 = censored

    study: study cohort